\documentclass[prl,noshowpacs,english,superscriptaddress,twocolumn]{revtex4-1}
\usepackage[colorlinks,linkcolor=blue,anchorcolor=blue,citecolor=blue,filecolor=blue,menucolor=blue,runcolor=blue,urlcolor=blue,frenchlinks=blue]{hyperref}
\usepackage{amsmath,amssymb,amsthm,amsfonts,mathrsfs}
\usepackage{graphicx}
\usepackage{pdfpages}
\usepackage{appendix}
\usepackage{lipsum}
\usepackage{slashed}
\usepackage{dcolumn}
\usepackage{mathrsfs}
\usepackage{bm}
\usepackage{color}
\usepackage{lineno,hyperref}
\makeatletter

\newcommand{\Rmnum}[1]{\expandafter\@slowromancap\romannumeral #1@}
\AtBeginDocument{\let\LS@rot\@undefined}
\makeatother
\usepackage{tikz,pgf}
\usetikzlibrary{shadings}
\usepackage{pgfplots,pgfplotstable}

\usepackage{tikz,pgf}
\usetikzlibrary{shadings}

\usepackage{pgfplots,pgfplotstable}

\newcommand{\GJ}{\textcolor{black}}
\begin{document}

\title{Floquet quantum geometry in periodically driven topological insulators}

\author{Peng He}
\email{hep@swjtu.edu.cn}
\affiliation{ School of Physical Science and Technology, Southwest Jiaotong University, Chengdu 610031, China}

\author{Jian-Te Wang}
\affiliation{Anhui Province Key Laboratory of Quantum Network, University of Science and Technology of China, Hefei 230026, China}
\affiliation{Synergetic Innovation Center of Quantum Information and Quantum Physics, University of Science and Technology of China, Hefei 230026, China}

\author{Jiangbin Gong}
\email{phygj@nus.edu.sg}
\affiliation{Department of Physics, National University of Singapore, Singapore 117551}
\affiliation{Centre for Quantum Technologies, National University of Singapore, Singapore 117543}
\affiliation{MajuLab, CNRS-UNS-NUS-NTU International Joint Research Unit, Singapore UMI 3654, Singapore}

\author{Hai-Tao Ding}
\email{htding.9@nus.edu.sg}
\affiliation{Centre for Quantum Technologies, National University of Singapore, Singapore 117543}
\affiliation{MajuLab, CNRS-UNS-NUS-NTU International Joint Research Unit, Singapore UMI 3654, Singapore}


\date{\today}

\begin{abstract}
\GJ{Quantum geometry plays a fundamental role across many branches of modern physics, yet its full characterization in nonequilibrium systems remains a challenge. Here, we propose a framework for quantum geometry in Floquet topological insulators by introducing a time-resolved quantum metric tensor, defined via the trace distance between micromotion operators in momentum-time space. For class A in two spatial dimensions,  we find a general inequality linking the Floquet quantum metric tensor and the Floquet topology: the associated quantum volume is bounded below by the Floquet topological invariant. This relation is found to also hold in class AIII in one dimension, where the Floquet geometric tensor may be notably reduced due to time-reflection symmetry.  This work will be useful in digesting the general aspects of quantum geometry in periodically driven systems in connection with their topological characterization. }

\end{abstract}

\maketitle

\textit{Introduction.}--Topology and geometry of quantum states in Hilbert space provide a fundamental framework for characterizing and classifying quantum matter \cite{Kolodrubetz2017,DWZhang2018,TLiu2024,JYurev2025}. While the topology of a quantum state manifold is described by a global invariant \cite{Hasan2010,XLQI2011}, its geometric properties are captured by the quantum geometric tensor, whose real part gives rise to \GJ{the} Fubini-Study metric, also called the quantum metric \cite{Provost1980,Berry1984}. Quantum metric plays essential roles in a plethora of physical phenomena related with nonadiabatic process, such as the nonlinear quantum Hall effect \cite{ZDu2021,AGa02023,NWang2023,JXLiu2025}, the phase coherence of topological superconductors \cite{Peotta2015,Julku2016,XHu2019,FXie2020,PHe2021,JXHu2025,ZCF2025,SAChen2024},  the optical conductivity \cite{Onishi2024}, and the orbital magnetic susceptibility \cite{YGao2015,Pechon2016}. \GJ{As a deep connection  between topology and geometry of quantum states, the topological invariant imposes a fundamental bound on the quantum metric \cite{CMJian2013,Ozawa2021,Mera2022,RRoyBand2014} in different symmetry classes }\cite{Palumbo2018,Zhu2021,AZhang2022,Ding2022,Jankowski2025,JYu2025}.  \GJ{Plus its experimental relevance,  the quantum metric serves as an excellent tool in the studies of topological quantum matter \cite{XTan2019,MYu2019,CRYi2023}.}

\GJ{With rapid progresses in diverse quantum platforms accommodating time-domain controls \cite{NHLindnerG2011,LJiang2011,Gong2012,Gong2013,Rechtsman2013, YHWangH2013,Asboth2014,Kundu2014,WZheng2014,Goldman2014,Eckardt2017,Potirniche2017,JWMcIverB2020,KWinterspergerC2020,SAfzalT2020,Huang2020,JYu2021,CChenXD2022}, topological physics in periodically driven (Floquet) systems is both rich and experimentally motivating}. \GJ{On the one hand, treating such Floquet systems typically starts with a time-independent effective  Hamiltonian yielding quasi-energy (or equivalently, Floquet eigenphase) bands \cite{Eckardt2017}. Because the quasi-energy is only defined up to modulus $2\pi$, Floquet systems in various symmetry classes \cite{Roy2017,SYao2017} yield both $0$ and $\pi$ gaps, leading to richer topological phases beyond the static counterparts \cite{Rudner2013,pireview}.  On the other hand, anomalous Floquet topological phases with zero band Chern number were predicted \cite{Rudner2013} and  experimentally verified in photonic systems \cite{Rechtsman2013,SAfzalT2020,CChenXD2022,Maczewsky2017,Mukherjee2017} and ultracold atoms \cite{KWinterspergerC2020,Wintersperger2020,JYZhang2023}.  Indeed, Floquet topological matter can be so anomalous such that an arbitrary number of chiral edge channels can still exist when the Floquet bands only admit a zero or unity Chern number \cite{LZhou2018G}.   This indicates that the link between geometry and topology of Floquet topological bands should be formulated  in a way different from static systems, covering both normal and anomalous cases.   }

In this Letter, \GJ{by treating momentum and time under the same footing,  we first propose a general framework to characterize the geometric structures of Floquet states. We define a Floquet quantum metric tensor (FQMT) by analyzing the trace distance between two neighbor micromotion operators within one driving period.  The FQMT thus defined captures the distinguishing nonequilibrium features of periodically driven systems within one driving period and does reduce to the conventional QMT in \textcolor{black}{static cases}.  We investigate the so-called Floquet quantum volume (FQV), namely, the integral of the square root of the FQMT determinant over the momentum-time space if no symmetry constraint is imposed. Remarkably, we prove that the FQV is bounded by the Floquet topological invariants from below, for both class A and class AIII. Theoretical results are verified and further illustrated by a Floquet Qi-Wu-Zhang model \cite{Bernevig2006,MCLiang2023,LZhang2020} and a Floquet Su-Schrieffer-Heeger (SSH) model as two specific examples \cite{WPSu1979,HWu2020}. Because signatures of Floquet band inversion are well manifested by the FQMT, the geometry and topology of Floquet quasienergy bands are hence linked closely, for both normal and anomalous Floquet phases}.  

\textit{Floquet geometry and topology in class A.}-- We consider a noninteracting lattice Hamiltonian with  periodic driving $H(\boldsymbol{k},t)=H(\boldsymbol{k},t+T)$ with period $T=2\pi/\omega$.  A complete set of basis of $H(\boldsymbol{k},t)$ determined by the quasienergy operator $Q(\boldsymbol{k},t)\equiv H(\boldsymbol{k},t)-i\partial_t $: $Q(\boldsymbol{k},t)|u_\alpha(\boldsymbol{k},t)\rangle=\mathcal{E}_\alpha|u_\alpha(\boldsymbol{k},t)\rangle$ where $|u_\alpha(\boldsymbol{k},t)\rangle$ and $\mathcal{E}_\alpha$ are called quasistationary states and quasienergies, respectively. Any given quasienergy $\mathcal{E}_\alpha$ can be associated with another one by $\mathcal{E}'_\alpha=\mathcal{E}_\alpha+n\omega$ with $n\in \mathbb{Z}$ due to the discrete time translational invariance. Therefore, the quasienergy can be defined within a regime dubbed quasienergy Brillouin zone $-\omega/2<\mathcal{E}_\alpha\le \omega/2$ \cite{Eckardt2017}. \GJ{A typical two-band Floquet system may be gapped at either $\mathcal{E}=0$ ( 0 gap) or $\mathcal{E}=\pi/T$ ($\pi$ gap), or both, an assumption made below.}

Accordingly, the time-evolution operator $U(\boldsymbol{k},t)=\mathcal{T}{\mathrm{ exp}}[-i\int_0^t H(\boldsymbol{k},\tau)d\tau]$ satisfies $U(\boldsymbol{k},t+T)=U(\boldsymbol{k},t)U(\boldsymbol{k},T)$. The topology and geometry of a Floquet system can be understood by decomposing $U(\boldsymbol{k},t)$ as a return map \cite{Rudner2013},
\begin{equation}
U(\boldsymbol{k}, t)=U_{\varepsilon}(\boldsymbol{k}, t)[U(\boldsymbol{k}, T)]_{\varepsilon}^{-t / T}.\label{eq_ut}
\end{equation}
with
\begin{equation}
[U(\boldsymbol{k}, T)]_{\varepsilon}^{-t / T}=\sum_\alpha \mathrm{exp}[-\frac{t}{T}{\mathrm {ln}}_\varepsilon e^{-i\mathcal{E}_\alpha T}] P_\alpha(\boldsymbol{k}, T) ,
\end{equation}
where $\alpha $ is the quasiband label, $P_\alpha \equiv |u_\alpha(\boldsymbol{k}, T)\rangle\langle u_\alpha(\boldsymbol{k}, T)|$ is the projection operator, $\varepsilon=0,\pi$ corresponds to different branch cuts such that the quasienergy $\mathcal{E}_{\alpha}$ in the $\alpha$-th band is chosen in the regime $[\varepsilon/T,(\varepsilon+2\pi)/T)$. Here, $[U(\boldsymbol{k}, T)]_{\varepsilon}$ describes an intuitive  part of $U(\boldsymbol{k}, t)$. Indeed, the one-period evolution operator $U(\boldsymbol{k}, T)$ allows us to define an effective Hamiltonian, $H_{\mathrm{eff}}=i\mathrm{ln}\left[U(\boldsymbol{k},T)\right]/T$, which captures the stroboscopic information. In contrast, the periodic
part $U_\varepsilon(\boldsymbol{k}, t)$ \GJ{accounts for Floquet topology from the micromotion of Floquet states within one driving period.}

To be concrete, we first consider periodically driven topological insulators in class A. Without loss of generality, we take the Chern insulator in two dimensions (2D) as an example. \textcolor{black}{In the case where only stroboscopic information is involved}, the system is well described by the effective Hamiltonian $H_{\mathrm{eff}}$. The topology is indexed as the conventional static version by the Chern number \cite{Thouless1982},
\begin{equation}
\mathcal{C}_\alpha=-\frac{1}{2 \pi} \int_{\mathbb{T}^2} d^2 \boldsymbol{k}~(\nabla_{\boldsymbol{k}} \times \mathcal{A}_\alpha ),
\end{equation}
where $\mathcal{A}_\alpha=\langle u_\alpha(\boldsymbol{k}) |i\nabla_{\boldsymbol{k}}|u_\alpha(\boldsymbol{k})\rangle$ is the Berry gauge field. On the other hand, the geometry of a band is characterized by the quantum metric tensor (see
 Supplemental Material (SM) for more details) \cite{sm},
\begin{equation}
g_{\mu \nu }^\alpha(\mathbf{k})=\frac{1}{2} \langle u^\alpha(\boldsymbol{k}) | \{r_{\mu  }, r_{\nu  } \} |u^\alpha(\boldsymbol{k}) \rangle .\label{eq_smt}
\end{equation}
where $\mu,\nu$ are spatial indices, and $\boldsymbol{r}=i\nabla_{\boldsymbol{k}}-\mathcal{A}_\alpha$ \cite{YPLin2022}. The quantum metric tensor quantifies the distance between two neighbor states in the Brillouin zone. $g_{\mu \nu }$ is symmetric $g_{\mu\nu}=g_{\nu\mu}$ and positive semidefinite $\mathrm{det}(g_{\mu\nu})\ge0$. We can further define the quantum volume as $\mathrm{vol}_{\mathrm{eff}}\equiv \int_{\mathbb{T}^2} d k_x d k_y \sqrt{\mathrm{det}(g_{\mu\nu})}$, which measures an effective size of the parameter space. Importantly, the quantum volume is bounded by the topological index \cite{Ozawa2021},
\begin{equation}
\mathrm{vol}_{\mathrm{eff}}\ge\pi|\mathcal{C}|.
\label{eqstatic}
\end{equation}
\GJ{Clearly, the relation Eq.~(\ref{eqstatic}) is obtained solely based on the time-independent effective Hamiltonian $H_{\rm eff}$, and as such, it cannot  capture the general relation between topology and geometry in Floquet systems where the micromotion within one driving period may dominate the physics. }

Indeed, \GJ{the Floquet band topology for class A,  at most partially captured by the Chern index associated with $H_{\rm eff}$, should be fully characterized by the following winding number} \cite{Rudner2013},
\begin{equation}
\mathcal{W}_{\varepsilon }=\int d t \frac{d^2 \boldsymbol{k}}{24 \pi^2} \epsilon_{i j k} \operatorname{Tr} (U_\varepsilon^{\dagger} \partial_i U_\varepsilon U_\varepsilon^{\dagger} \partial_j U_\varepsilon U_\varepsilon^{\dagger} \partial_k U_\varepsilon ).\label{eq_fwn}
\end{equation}  \GJ{Unlike the Floquet band Chern number, the winding number $\mathcal{W}_{\varepsilon }$  can be used to fully determine the number of Floquet edge states. 
It is thus necessary and highly useful to explicitly connect the topological index $\mathcal{W}_{\varepsilon }$ with the geometrical aspects of Floquet states. }

\begin{figure}[htbp]
	\centering
	\includegraphics[width=0.48\textwidth]{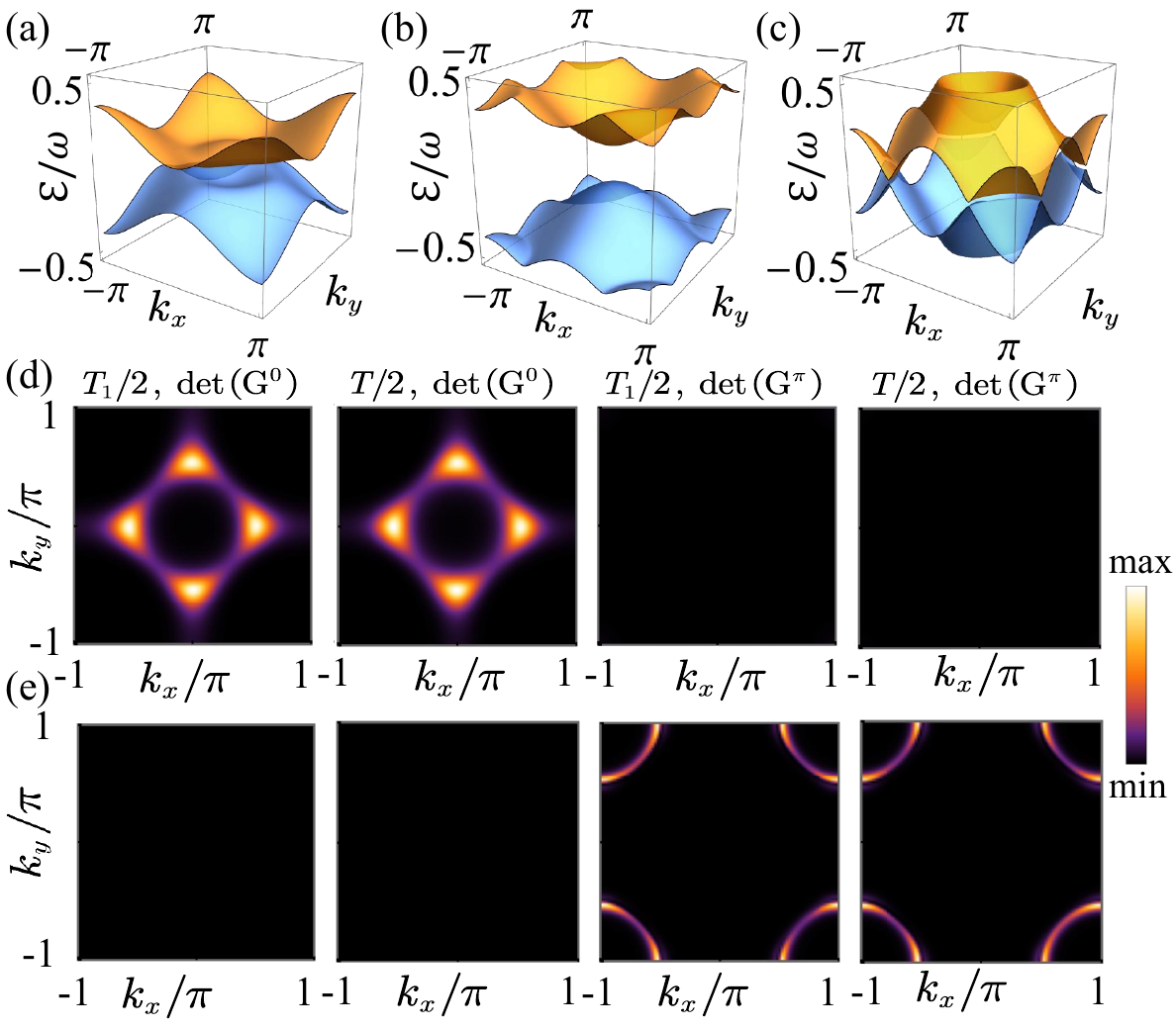}
	\caption{(a)-(c) Quasienergy band structures of the Floquet  Qi-Wu-Zhang model in the first Brillouin zone in the topological 0 phase (a), $\pi$ phase (b) and $0$-$\pi$ phase (c). (d)-(e) The snapshots of the determinants of the FQMT at different times, with (d) in the topological 0 phase and (e) in the topological $\pi$ phase with the same system parameters as in (a) and (b). Here $v_x=v_y=v_z=1$ are used as energy units and the mass terms is (a) $m=5$, (b) $m=15$ and (c) $m=15$. The driving frequency is set as (a) $\omega=8$, (b) $\omega=8$ and (c) $\omega=4$, and $T_1=0.6T$, $T_2=0.4T$.}
	\label{fig1}
\end{figure}

We are hence motivated to define a Floquet quantum metric tensor in the momentum-time continuum $\boldsymbol{p}=(\boldsymbol{k},t)$, instead of using the momentum variable only for the quantum geometry based on Floquet eigenstates of the effective Hamiltonian~\cite{LZhou2024}.
Consider then the trace distance between $U_\varepsilon(\boldsymbol{p})$ and $U_\varepsilon(\boldsymbol{p}+\delta \boldsymbol{p})$, which reads $D=\frac{1}{2}\operatorname{Tr}\sqrt{\delta U_\varepsilon(\boldsymbol{p})^\dagger \delta U_\varepsilon(\boldsymbol{p})}$ with $\delta U_\varepsilon(\boldsymbol{p})=U_\varepsilon(\boldsymbol{p}+\delta \boldsymbol{p})-U_\varepsilon(\boldsymbol{p})$ \cite{Nielsen2010}. Naturally,  a Floquet quantum metric tensor can be defined as the following: $dl^2=2D^2=\sum_{\mu \nu} G_{\mu\nu}dp^\mu dp^\nu$. By expanding $U_\varepsilon(\boldsymbol{p}+\delta \boldsymbol{p})$ to second order (see SM for computation details) \cite{sm}, we have
\begin{equation}
G^\varepsilon_{\mu \nu}(\boldsymbol{p})=\frac{1}{2}\operatorname{Tr}[\partial_{p_\mu}U^\dagger_\varepsilon(\boldsymbol{p}) \partial_{p_\nu}U_\varepsilon(\boldsymbol{p})],
\end{equation}
where $G^\varepsilon_{\mu \nu}$ is a $3\times 3$ real symmetric tensor defined in the space $p_\mu,p_\nu=\{p_x,p_y,t\}$. $G^{\varepsilon=0(\pi)}_{\mu \nu}$ accounts for the quantum geometry around the Floquet phase gap $0$ ($\pi$). Similarly, we can also define the Floquet quantum volume,
\begin{equation}
 \mathrm{vol}_\varepsilon \equiv \int d^3 \boldsymbol{p} \sqrt{\mathrm{det}(G^\varepsilon_{\mu\nu})},
\end{equation}
As one central result of this work, we are able to find that the Floquet quantum volume defined above is also bounded by the Floquet topology from below, namely, 
\begin{equation}
\mathrm{vol}_\varepsilon \ge 2\pi^2 |W_{\varepsilon }|.\label{eq_bd}
\end{equation}
\GJ{The proof of this general relation between the quantum metric and the topological invariant in nonequilibrium systems is done} by considering a minimal two-band model. We decompose the Floquet operator as $U_\varepsilon=d^\varepsilon_0\sigma_0-i\sum_{i=1}^3 d_i^\varepsilon\sigma_i$ with $\sigma_0=\mathbb{I}_2$ being the identity matrix and $\sigma_{i=x,y,z}$ being the Pauli matrices. Then we reduce the expression of the winding number to $W_\varepsilon=\frac{1}{2 \pi^2} \int_{\mathbb{T}^3} d^3 \boldsymbol{p} \epsilon^{\mu \nu \rho \lambda} d^\varepsilon_{ \mu }\partial_{k_x} d^\varepsilon_{\nu} \partial_{k_y} d^\varepsilon_{ \rho} \partial_{t} d^\varepsilon_{ \lambda}$, as well as the FQM $G^\varepsilon_{\mu \nu}=\sum_{i=0}^3\partial_{p_\mu} d_i^ \varepsilon \partial_{p_\nu} d_i^ \varepsilon$. Then by proving that $\mathrm{det}(G^\varepsilon)^2=| \epsilon^{\mu \nu \rho \lambda} d^\varepsilon_{ \mu }\partial_{k_x} d^\varepsilon_{\nu} \partial_{k_y} d^\varepsilon_{ \rho} \partial_{t} d^\varepsilon_{ \lambda}|$ (see SM for detailed proof) \cite{sm}, we arrive at Eq. (\ref{eq_bd}).

To verify our results, we now take a 2D Floquet Qi-Wu-Zhang model as a working example, which has been experimentally realized with ultracold atoms in optical Raman lattices recently \cite{JYZhang2023}. The Bloch Hamiltonian is given by $H_1(\boldsymbol{k},t)=\boldsymbol{h}(\boldsymbol{k},t)\cdot\boldsymbol{\sigma}$, with $\boldsymbol{h}(\boldsymbol{k},t)=(v_x\sin k_x,v_y\sin k_y,M(t)-v_z\cos k_x-v_z\cos k_y)$ and $v_{x,y,z}$ denoting the Fermi velocities. The mass term $M(t)$ is under a step driving \cite{LZhang2020},
\begin{equation}
M(t)= \{\begin{array}{ll}
m, & t \in [n T, n T+T_1 ) \\
-m, & t \in [n T+T_1,(n+1) T ),
\end{array} \quad n \in \mathbb{Z}, 
\end{equation}
and for convenience we denote $T_2=T-T_1$ below. This model features three different topological phases: the topological 0 phase \GJ{($\mathcal{W}_{0}=\pm 1$, $\mathcal{W}_{\pi}=0$)} , $\pi$ phase \GJ{($\mathcal{W}_{\pi}=\pm 1$, $\mathcal{W}_{0}=0$}), and $0$-$\pi$ phase \GJ{($\mathcal{W}_{0}=\pm 1$, $\mathcal{W}_{\pi}=\pm 1$)}, with topological edge states residing in the corresponding quasienergy gaps. In Fig.~\ref{fig1}(a)-\ref{fig1}(c), we plot the typical Floquet quasienergy bands in each phase with periodic boundary conditions. The Floquet quasienergy bands have band inversions around the gaps centered at $0$ and $\pi$ due to their topological nature \cite{sm}. The values of the FQMT \GJ{at different times} are shown in Figs.~\ref{fig1}(d) and \ref{fig1}(e), in terms of $\mathrm{det}({G^\varepsilon})$. It is seen that $\mathrm{det}({G^\varepsilon})$ acquires prominent values near the band inversion lines in the topological phase, but not in the topological trivial case. Furthermore, $\mathrm{det}({G^0})$ is seen to be featureless in the $\pi$ phase, whereas $\mathrm{det}({G^\pi})$ has no feature in the 0 phase.   \GJ{These results clearly indicate that the FQMT proposed here is a useful diagnosis tool to study Floquet band topology.}

\begin{figure}[htbp]
	\centering
	\includegraphics[width=0.48\textwidth]{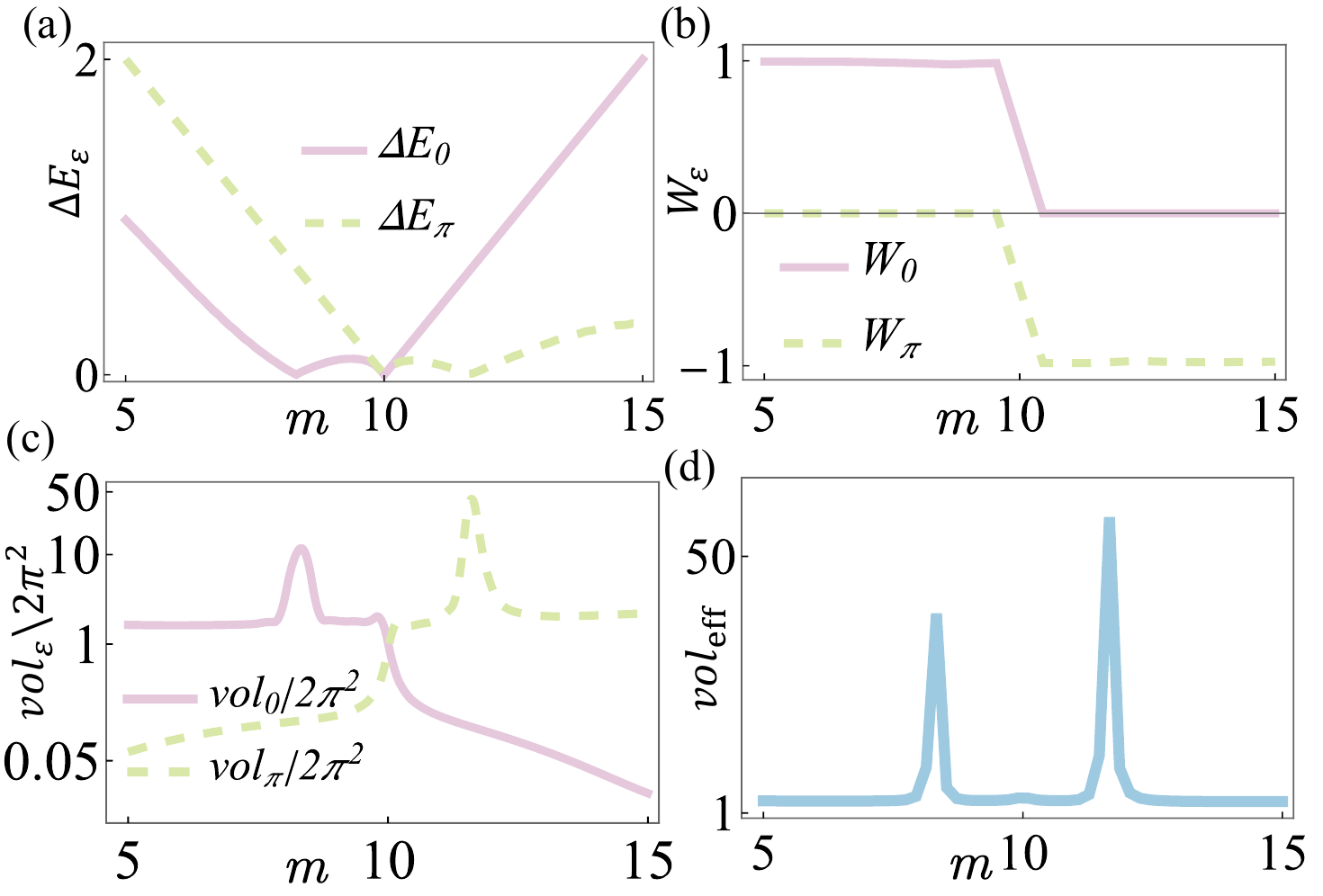}
	\caption{(a) The quasienergy gap for Floquet  Qi-Wu-Zhang model as the function of mass term $m$. (b) The winding numbers $W_0$ and $W_\pi$ as the function of mass term $m$. (c) The Floquet quantum volume $\mathrm{vol}_0/(2\pi^2)$ and $\mathrm{vol}_\pi/(2\pi^2)$ as the function of mass term $m$. (d)  The quantum volume of the effective Hamiltonian  $\mathrm{vol}_{\mathrm{eff}}$ as the function of mass term $m$. Here $v_x=v_y=v_z=1$, $\omega=8$, and $T_1=0.6T$, $T_2=0.4T$. }
	\label{fig2}
\end{figure}

\GJ{To further digest the usefulness of FQMT, we examine its behavior in Fig.~\ref{fig2} as one system parameter is tuned continuously. Fig.~\ref{fig2}(a) depicts how the Floquet bands centered at $0$ or $\pi$ gap close twice, triggering topological phase transitions, as confirmed by the topological winding numbers shown in Fig.~\ref{fig2}(b).  Echoing with the observations made in Fig.~1, Fig.~\ref{fig2}(c) shows that the FQV obtained from the FQMT is much larger in the topological regime than in the topological trivial regime.  By contrast, the quantum volume shown in Fig.~\ref{fig2}(d) (predicted by the effective Hamiltonian $H_{\text{eff}}$) shows rather uniform behaviors, except for some divergence behavior at gap closing points (divergence in the normal quantum metric for $H_{\text{eff}}$ was also reported in Ref.~\cite{LZhou2024}). Such QV hence cannot distinguish between the 0 phase and $\pi$ phases.  }

Some remarks are \GJ{in order}. \GJ{Firstly, the inequality of Eq.~(\ref{eq_bd}) can actually be saturated by a special class of topological insulators whose sign of the integrand in Eq.~(\ref{eq_fwn}) are uniform over the entire BZ.  Secondly, the FQMT $G_{k_x,k_y}$ reduces to the conventional QMT $g_{k_x,k_y}$ in the high-frequency limit \cite{sm}. Finally, for an anomalous Floquet phase with $\mathcal{W}_0=\mathcal{W}_\pi=1$ and hence with a zero Chern number $\mathcal{C}={\mathcal W}_0-\mathcal{W}_\pi=0$,  $\mathrm{vol}_{\mathrm{eff}}$ has a trivial low bound whereas both $\mathrm{vol}_0$ and $\mathrm{vol}_\pi$ have nontrivial lower bounds}.  

\begin{figure}[htbp]
	\centering
	\includegraphics[width=0.48\textwidth]{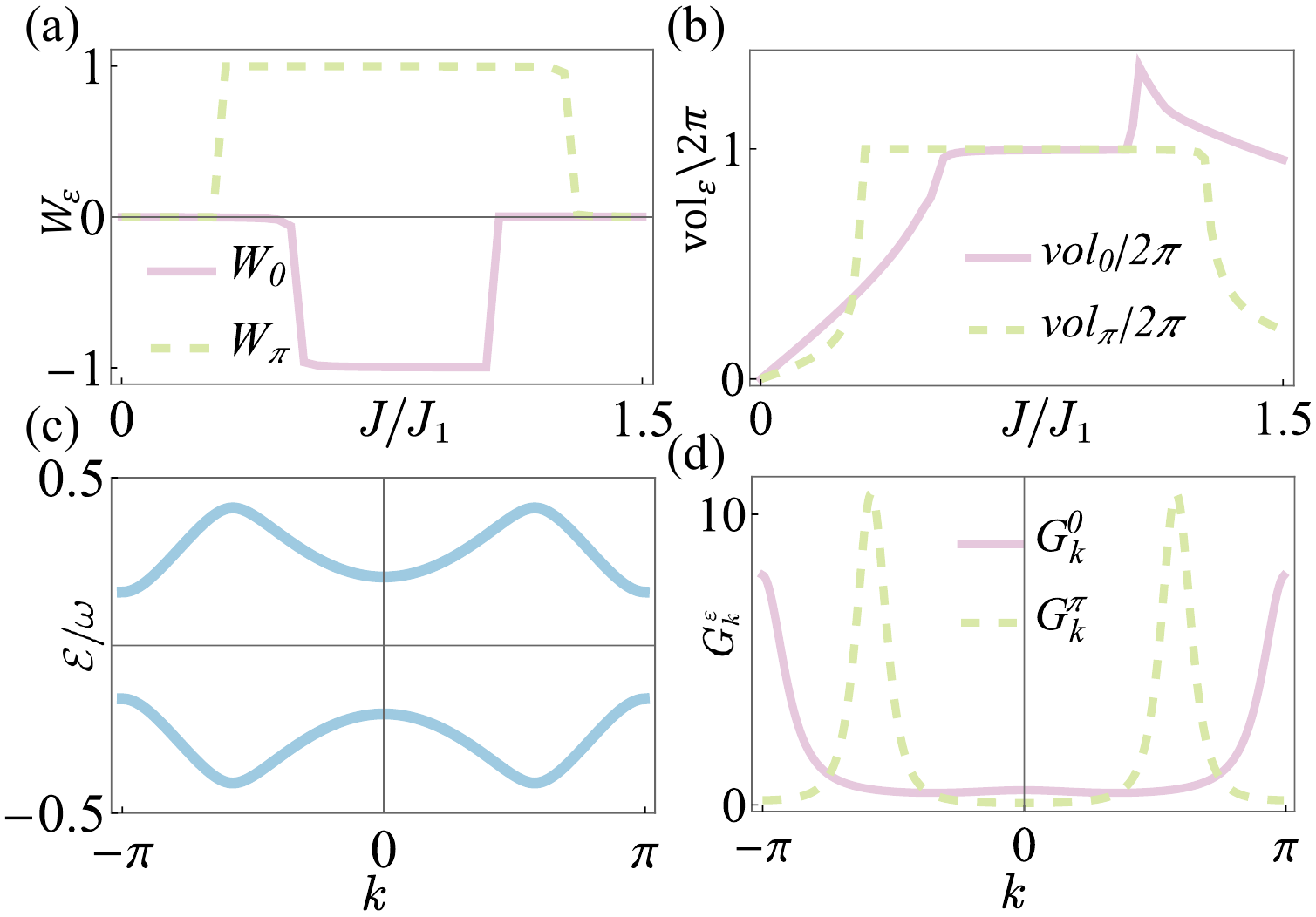}
	\caption{(a) The winding numbers $W_0$ and $W_\pi$ for the Floquet SSH model vs the system parameter $J/J_1$. (b) The FQV $\mathrm{vol}_0/(2\pi)$ and $\mathrm{vol}_\pi/(2\pi)$ vs $J/J_1$. (c) A typical quasienergy spectrum in the topological 0-$\pi$ phase with $J/J_1=0.75$. (d) The FQV corresponding to (c). Here other system parameters are chosen to be $q=3$, $\omega=2\pi$, $T_1=T_2$.}
	\label{fig3}
\end{figure}

\textit{Reduction of Floquet geometric tensor in class AIII.}-- \GJ{In the high-frequency driving case,  the physics of Floquet band topology reduces to that for the one-period effective Hamiltonian $H_{\rm eff}$. As shown in SM~\cite{sm}, in this case the FQMT reduces to the familiar quantum metric in static systems. Here we show that our general treatment of the FQMT may be also reduced in other scenarios, using the topological class AIII in one dimension (1D) as one example.  Consider then the following 1D case, with the Floquet operator respecting the chiral symmetry \cite{Roy2017,SYao2017},
\begin{equation}
S^{-1}U_\varepsilon(k,t) S=U_{-\varepsilon}(k,-t)\mathrm{exp}(i\frac{2\pi t}{T}),\label{eq_cs}
\end{equation}
where $S$ is the \GJ{chiral} symmetry operator. \textcolor{black}{Eq.~(\ref{eq_cs}) shows that the chiral symmetry in Floquet systems acts as a time-reflection symmetry, which leads to $S^{-1}U_\varepsilon(k,T/2) S=\mp U_\varepsilon(k,T/2)$, with $\mp$ for $\varepsilon=0,\pi$.} The associated topological invariant was previously found to be
\begin{equation}
W_\varepsilon=\int \frac{dk}{4\pi}\operatorname{Tr}[SU_\varepsilon^\dagger(k,T/2)i\partial_kU_\varepsilon(k,T/2)].
\end{equation}}
\GJ{Interestingly, here only the time evolution operator $U_\varepsilon$ at fixed $t=T/2$ suffices to characterize the Floquet topology. This is made  possible by symmetry, together with the fact that $U_\varepsilon$ captures not only the state features at $T/2$, but also other features such as its one-period Floquet operator $U(T)$ \cite{Fruchart2016}.} 

\GJ{Focusing on $U_\varepsilon(k,T/2)$ at fixed time $t=T/2$, We are hence left with only the $k$ variable for 1D systems. Our general treatment above then takes us to the following drastically reduced FQMT with one continuous index $k$, namely,}
\begin{equation}
G^\varepsilon_k=\frac{1}{2}\operatorname{Tr}[\partial_k U_\varepsilon^\dagger(k,T/2) \partial_k U_\varepsilon(k,T/2)].
\end{equation}
Curiously,  even in this reduced form, the FQV for this chiral symmetric class, namely, $\mathrm{vol}_\varepsilon=\int dk \sqrt{G^\varepsilon_k}$,
is still bounded by the topological winding number here, i.e.,  $\mathrm{vol}_\varepsilon\ge 2\pi|W_\varepsilon|$.  This result hence strengthens the connection between the FQV and the topological characterization of nonequilibrium topology.    

To verify our results here, we turn to a Floquet Su-Schrieffer–Heeger (SSH) model, with the model Hamiltonian $H_2(k)=(J_1+J_2
\cos k)\sigma_x+J_2\sin k\sigma_y$, where $J_1$ denotes the intra-cell hopping amplitude and $J_2$ the inter-cell hopping amplitude \cite{Fruchart2016}. The inter-cell hopping amplitude is under a symmetric step driving,
\begin{equation}
J_2(t)= \left\{\begin{array}{lll}
J, & t \in [n T, n T+T_1/2 ) \\
qJ, & t \in [n T+T_1/2,n T+T_1/2+T_2)\\
J, & t \in [n T+T_1/2+T_2,(n+1) T ),
\end{array}   \right.
\end{equation}
where $T=T_1+T_2$, and $q$ is a real-valued coefficient.
Figs.~\ref{fig3}(a) and  \ref{fig3}(b) present the phase diagram and the corresponding FQV. Figure~\ref{fig3}(c) depicts a typical Floquet band structure.  The FQMT over the BZ in the topological 0-$\pi$ phase is shown in Fig.~\ref{fig3}(d). \GJ{It is seen that $G^0_k$ and $G^\pi_k$ peak at the band minimum or band maximum where the band inversions happen around the 0 gap and $\pi$ gap respectively. The topological bound of the FQV is verified in Fig.~\ref{fig3}(b).  One may also explore other system parameter regimes where the Floquet SSH model may acquire larger winding numbers. In these cases, as shown in the SM~\cite{sm}, larger values of FQV are obtained, thus illustrating again the usefulness of the FQV to manifest topological phase transitions.   Importantly, even if we retain the full momentum-time structure of the FQMT without really using the time-reflection symmetry for reduction,  the full FQMT in the momentum-time space is still useful in analyzing topological phase transitions, with details also shown in the SM~\cite{sm}.}

\textit{Detection scheme.}-- For static systems, the QMT defined in Eq. (\ref{eq_smt}) can be extracted via the excitation
 rate following a sudden quench \cite{LKLim2015}, or the Rabi oscillations upon parametric modulations \cite{Ozawa2018}, both of which have been experimentally demonstrated in state-of-the-art quantum  simulation platforms \cite{XTan2019,MYu2019,CRYi2023}. For periodically driven quantum systems, there are several efficient schemes to probe the winding numbers, by converting it to the numbers of singularities owing to its topological nature \cite{LZhang2020,Unal2019}. Here, we propose a direct probe method for the FQMT, based on the full tomography of the
evolution operator $U(\boldsymbol{k},t)$.

For the two band model used in this paper, the
evolution operator $U(\boldsymbol{k},t)$ are parameterized as 
\begin{equation}
U=\cos \chi_{\boldsymbol{k}}t-i \sin \chi_{\boldsymbol{k}}t \left[\begin{array}{cc}
\cos \theta_{\boldsymbol{k}} & \sin \theta_{\boldsymbol{k}} e^{-i \varphi_{\boldsymbol{k}}} \\
\sin \theta_{\boldsymbol{k}} e^{i \varphi_{\boldsymbol{k}}} & -\cos \theta_{\boldsymbol{k}}
\end{array} \right],
\end{equation}
with three angles $\chi_{\boldsymbol{k}},\theta_{\boldsymbol{k}}\in [0,\pi]$ and $\varphi_{\boldsymbol{k}}\in (-\pi,\pi]$ \cite{Huang2020}. By preparing the system to its Floquet eigenstate $\Psi_{\boldsymbol{k}}=(\cos \theta_{\boldsymbol{k}}/2,e^{i \varphi_{\boldsymbol{k}}}\sin \theta_{\boldsymbol{k}}/2)^{\mathrm{T}}$, then turning off the Hamiltonian and switching on a Zeeman field $H'=\Delta/2\sigma_z$, the angles $\theta_{\boldsymbol{k}}$ and $\varphi_{\boldsymbol{k}}$ can be obtained by measuring the dynamics of momentum space population $n_{\boldsymbol{k}}=1-\sin \theta_{\boldsymbol{k}}\cos(\varphi_{\boldsymbol{k}}+\Delta t)$ \cite{Hauke2014}. Such a quench protocol is feasible for current technologies of ultracold atoms \cite{Tarnowski2019}. Furthermore, $\chi_{\boldsymbol{k}}$ can be extracted from the absorption spectroscopy with coupling to auxiliary degrees of freedom \cite{Vale2021,QLiang2022,PHe2025}.
With the full tomography of $U(\boldsymbol{k},t)$ executed, we can further reconstruct the micromotion operator by a smooth return map $U\mapsto U_\varepsilon$ that leaves the gap $\varepsilon$ open~\cite{Unal2019}. With all these steps,  one can indeed extract the FQMT via the pixelized $U_\varepsilon(\boldsymbol{k},t)$. Although the topological invariants and the FQMT are defined in terms of $U_\varepsilon(t)$, we can still rewrite them using the evolved state from arbitrary initial state (for details, see SM) \cite{sm}, then the winding number will reduce to a Hopf number, which also facilitates its measurement \cite{UnalHf2019}.

\textit{Concluding remarks}--
\GJ{Motivated by the unusual richness of nonequilibrium topological phases of matter, we have proposed a meaningful treatment to characterize the geometry of Floquet topological insulators, shedding light on the close connection between the geometry of Floquet systems and their topological characterization.  Though mainly working on systems in 2D in class A where momentum and time variables are treated under the same footing, we have also used class A\uppercase\expandafter{\romannumeral3} in 1D to understand how a reduction in the Floquet quantum metric tensor may be possible if one fully uses the symmetry properties of the time evolution operator.
As possible future work, it should be fruitful to extend our results to 4-dimensional topological insulators with a nontrivial second Chern number \cite{Price2015,Zhu2022} as well as 3-dimensional chiral insulators \cite{STWang2014}. It also remains to explore gapped system in other symmetry classes and gapless systems including Floquet topological semimetals \cite{GongPRLsemimetal}.}
 
~
\acknowledgments
We thank H. Wu for helpful discussions.  J.G. acknowledges support by the National Research Foundation, Singapore, through the National Quantum Office, hosted in A*STAR, under its Centre for Quantum Technologies Funding Initiative (S24Q2d0009).


\begin{appendix}

\end{appendix}

\widetext
\clearpage
	


\section*{Supplementary material}

# Supplemental Materials for "Floquet quantum geometry in periodically driven topological insulators"

Peng He,[1, *] Jian-Te Wang,[2, 3] Jiangbin Gong,[4, 5, †] and Hai-Tao Ding[5, 6, ‡]

[1]*School of Physical Science and Technology, Southwest Jiaotong University, Chengdu 610031, China*
[2]*Anhui Province Key Laboratory of Quantum Network,*
*University of Science and Technology of China, Hefei 230026, China*
[3]*Synergetic Innovation Center of Quantum Information and Quantum Physics,*
*University of Science and Technology of China, Hefei 230026, China*
[4]*Department of Physics, National University of Singapore, Singapore 117551*
[5]*Centre for Quantum Technologies, National University of Singapore, Singapore 117543*
[6]*MajuLab, CNRS-UNS-NUS-NTU International Joint Research Unit, Singapore UMI 3654, Singapore*

## CONTENTS

## S-I. PRELIMINARIES OF QUANTUM METRIC TENSOR IN STATIC SYSTEMS

For a nondegenerate ground state manifold $\{|\psi(\boldsymbol{k})\rangle\}$ in D-dimensional Brillouin zone $\boldsymbol{k} = (k_1, k_2, \ldots, k_D)$, the distance between two neighboring states $|\psi(\boldsymbol{k})\rangle$ and $|\psi(\boldsymbol{k} + \delta\boldsymbol{k})\rangle$ is given by [1],

$$ds^2 = 1 - |\langle\psi(\boldsymbol{k})|\psi(\boldsymbol{k} + \delta\boldsymbol{k})\rangle|^2 = \sum_{\mu\nu} g_{\mu\nu} dk_\mu dk_\nu. \tag{S1}$$

Here $g_{\mu\nu}$ is the quantum metric tensor (QMT),

$$g_{\mu\nu} = \mathrm{Re}[\langle\partial_{k_\mu}\psi(\boldsymbol{k})|\partial_{k_\nu}\psi(\boldsymbol{k})\rangle] - \langle\partial_{k_\mu}\psi(\boldsymbol{k})|\psi(\boldsymbol{k})\rangle\langle\psi(\boldsymbol{k})|\partial_{k_\nu}\psi(\boldsymbol{k})\rangle. \tag{S2}$$

In fact, the quantum metric tensor and the Berry curvature can be unified in the framework of the quantum geometric tensor (QGT),

$$\mathcal{G}_{\mu\nu} = \langle\partial_{k_\mu}\psi(\boldsymbol{k})|[1 - |\psi(\boldsymbol{k})\rangle\langle\psi(\boldsymbol{k})|]|\partial_{k_\nu}\psi(\boldsymbol{k})\rangle, \tag{S3}$$

———————
* hep@swjtu.edu.cn
† phygj@nus.edu.sg
‡ htding.9@nus.edu.sg

where the real part of QGT is the QMT: $\mathrm{Re}(\mathcal{G}_{\mu\nu}) = g_{\mu\nu}$, whereas the imaginary part $\mathrm{Im}(\mathcal{G}_{\mu\nu}) = -i(\langle \partial_{k_\mu}\psi(\boldsymbol{k})|\partial_{k_\nu}\psi(\boldsymbol{k})\rangle - \langle \partial_{k_\nu}\psi(\boldsymbol{k})|\partial_{k_\mu}\psi(\boldsymbol{k})\rangle)/2$ is exactly the Berry curvature.

We can also write down the QGT and QMT in terms of the projection operator $P(\boldsymbol{k}) \equiv |\psi(\boldsymbol{k})\rangle\langle\psi(\boldsymbol{k})|$,

$$\mathcal{G}_{\mu\nu} = \mathrm{Tr}[P(\boldsymbol{k})\partial_{k_\mu}P(\boldsymbol{k})\partial_{k_\nu}P(\boldsymbol{k})], \tag{S4}$$

$$g_{\mu\nu} = \mathrm{Tr}[\partial_{k_\mu}P(\boldsymbol{k})\partial_{k_\nu}P(\boldsymbol{k})]. \tag{S5}$$

Specifically, for a two-band model with a Bloch Hamiltonian $H(\boldsymbol{k}) = \sum_i d_i\sigma_i$ ($d_i$ is real-valued and the Hamiltonian is flattened such that $\sum_i d_i^2 = 1$), the projection operator for the ground state is $P(\boldsymbol{k}) = \frac{1}{2}(1 - \sum_i d_i\sigma_i)$. Then the QMT is simply $g_{\mu\nu} = \frac{1}{2}(\partial_{k_\mu}\boldsymbol{d}) \cdot (\partial_{k_\nu}\boldsymbol{d})$, with $\boldsymbol{d} = (d_1, d_2, d_3)$.

On the other hand, using the eigenequation of the Hamiltonian $H(\boldsymbol{k})|\psi_m(\boldsymbol{k})\rangle = E_m(\boldsymbol{k})|\psi_m(\boldsymbol{k})\rangle$, we can rewrite the QGT as

$$\mathcal{G}_{\mu\nu} = \sum_{m=1}^{n}\sum_{m'>n} \frac{\langle\psi_m(\boldsymbol{k})|\partial_{k_\mu}H|\psi_{m'}(\boldsymbol{k})\rangle\langle\psi_{m'}(\boldsymbol{k})|\partial_{k_\nu}H|\psi_m(\boldsymbol{k})\rangle}{(E_{m'}(\boldsymbol{k}) - E_m(\boldsymbol{k}))^2}, \tag{S6}$$

where $n$ is the number of occupied states. Here relation $\langle\psi_{m'}(\boldsymbol{k})|\partial_{k_\mu}H|\psi_m(\boldsymbol{k})\rangle = (E_m(\boldsymbol{k}) - E_{m'}(\boldsymbol{k}))\langle\psi_{m'}(\boldsymbol{k})|\partial_{k_\nu}\psi_m(\boldsymbol{k})\rangle$ is used. From Eq. (S6), we can see that the QGT would diverge at the gap closing point.

## S-II. DEFINITION OF FLOQUET QUANTUM METRIC TENSOR

Given two operators $\rho$ and $\sigma$, their trace distance is given by $D(\rho, \sigma) = \frac{1}{2}\mathrm{Tr}[\sqrt{(\rho - \sigma)^\dagger(\rho - \sigma)}]$. On the other hand, one can also define the fidelity between them as $F(\rho, \sigma) = \mathrm{Tr}[\rho^{1/2}\sigma\rho^{1/2}]$ [2]. The trace distance is related to the fidelity through $D(\rho, \sigma) = \sqrt{1 - F^2(\rho, \sigma)}$, if $\rho$ and $\sigma$ are density matrices for pure states. Now we consider the trace distance between two time evolution operators (defined in the main text) $U_\varepsilon(\boldsymbol{p})$ and $U_\varepsilon(\boldsymbol{p}+\delta\boldsymbol{p})$ in the momentum-time space $\boldsymbol{p} = (\boldsymbol{k}, t)$,

$$D = \frac{1}{2}\mathrm{Tr}\sqrt{[U_\varepsilon(\boldsymbol{p}+\delta\boldsymbol{p}) - U_\varepsilon(\boldsymbol{p})]^\dagger[U_\varepsilon(\boldsymbol{p}+\delta\boldsymbol{p}) - U_\varepsilon(\boldsymbol{p})]}. \tag{S7}$$

Then we have

$$2D^2 = \frac{1}{2}\mathrm{Tr}[U_\varepsilon^\dagger(\boldsymbol{p}+\delta\boldsymbol{p})U_\varepsilon(\boldsymbol{p}+\delta\boldsymbol{p}) - U_\varepsilon^\dagger(\boldsymbol{p}+\delta\boldsymbol{p})U_\varepsilon(\boldsymbol{p}) - U_\varepsilon^\dagger(\boldsymbol{p})U_\varepsilon(\boldsymbol{p}+\delta\boldsymbol{p}) + U_\varepsilon^\dagger(\boldsymbol{p})U_\varepsilon(\boldsymbol{p})]. \tag{S8}$$

The Floquet quantum metric tensor (FQMT) is given by $dl^2 = 2D^2 = G_{\mu\nu}dp^\mu dp^\nu$. By expanding $U_\varepsilon(\boldsymbol{p}+\delta\boldsymbol{p})$ up to the second order,

$$U_\varepsilon(\boldsymbol{p}+\delta\boldsymbol{p}) = U_\varepsilon(\boldsymbol{p}) + \sum_\mu \partial_{p_\mu}U_\varepsilon(\boldsymbol{p})dp_\mu + \frac{1}{2}\sum_{\mu,\nu}\partial_{p_\mu}\partial_{p_\nu}U_\varepsilon(\boldsymbol{p})dp_\mu dp_\nu. \tag{S9}$$

we have

$$\begin{aligned} U_\varepsilon^\dagger(\boldsymbol{p}+\delta\boldsymbol{p})U_\varepsilon(\boldsymbol{p}+\delta\boldsymbol{p}) =& U_\varepsilon^\dagger(\boldsymbol{p})U_\varepsilon(\boldsymbol{p}) + \sum_\nu U_\varepsilon^\dagger(\boldsymbol{p})\partial_{p_\nu}U_\varepsilon(\boldsymbol{p})dp_\nu + \sum_{\mu,\nu}\frac{1}{2}U_\varepsilon^\dagger(\boldsymbol{p})\partial_{\boldsymbol{p}_\mu}\partial_{p_\nu}U_\varepsilon(\boldsymbol{p})dp_\mu dp_\nu \\ &+ \sum_\mu \partial_{p_\mu}U_\varepsilon^\dagger(\boldsymbol{p})U_\varepsilon(\boldsymbol{p})dp_\mu + \sum_{\mu,\nu}\partial_{p_\mu}U_\varepsilon^\dagger(\boldsymbol{p})\partial_{p_\nu}U_\varepsilon(\boldsymbol{p})dp_\mu dp_\nu + \sum_{\mu,\nu}\frac{1}{2}\partial_{p_\mu}\partial_{p_\nu}U_\varepsilon^\dagger(\boldsymbol{p})U_\varepsilon(\boldsymbol{p})dp_\mu dp_\nu, \end{aligned} \tag{S10}$$

$$U_\varepsilon^\dagger(\boldsymbol{p}+\delta\boldsymbol{p})U_\varepsilon(\boldsymbol{p}) = U_\varepsilon^\dagger(\boldsymbol{p})U_\varepsilon(\boldsymbol{p}) + \sum_\mu \partial_{p_\mu}U_\varepsilon^\dagger(\boldsymbol{p})U_\varepsilon(\boldsymbol{p})dp_\mu + \sum_{\mu,\nu}\frac{1}{2}\partial_{\boldsymbol{p}_\mu}\partial_{p_\nu}U_\varepsilon^\dagger(\boldsymbol{p})U_\varepsilon(\boldsymbol{p})dp_\mu d\boldsymbol{p}_\nu, \tag{S11}$$

$$U_\varepsilon^\dagger(\boldsymbol{p})U_\varepsilon(\boldsymbol{p}+\delta\boldsymbol{p}) = U_\varepsilon^\dagger(\boldsymbol{p})U_\varepsilon(\boldsymbol{p}) + \sum_\mu U_\varepsilon^\dagger(\boldsymbol{p})\partial_{p_\mu}U_\varepsilon(\boldsymbol{p})dp_\mu + \sum_{\mu,\nu}\frac{1}{2}U_\varepsilon^\dagger(\boldsymbol{p})\partial_{p_\mu}\partial_{p_\nu}U_\varepsilon(\boldsymbol{p})dp_\mu dp_\nu. \tag{S12}$$

Substituting Eqs. (S10)-(S12) into Eq. (S9), we have

$$2D^2 = \sum_{\mu,\nu} \frac{1}{2} \text{Tr}[\partial_{p_\mu} U_\varepsilon^\dagger(\boldsymbol{p}) \partial_{p_\nu} U_\varepsilon(\boldsymbol{p})] dp_\mu dp_\nu,$$ (S13)

which yields our definition of the FQMT $G_{\mu\nu}^\varepsilon$ for $U_\varepsilon(\boldsymbol{p})$, namely,

$$G_{\mu\nu}^\varepsilon(\boldsymbol{p}) = \frac{1}{2} \text{Tr} \left[ \partial_{p_\mu} U_\varepsilon^\dagger(\boldsymbol{p}) \partial_{p_\nu} U_\varepsilon(\boldsymbol{p}) \right].$$ (S14)

The FQMT $G_{\mu\nu}^\varepsilon$ is real and symmetric, $G_{\mu\nu}^\varepsilon = G_{\nu\mu}^\varepsilon = (G_{\mu\nu}^\varepsilon)^* = (G_{\nu\mu}^\varepsilon)^*$. Using the unitarity of $U_\varepsilon(\boldsymbol{p})$, $U_\varepsilon^\dagger(\boldsymbol{p}) U_\varepsilon(\boldsymbol{p}) = \mathbb{I}$, we have $\partial_{p_\mu} U_\varepsilon^\dagger(\boldsymbol{p}) U_\varepsilon(\boldsymbol{p}) + U_\varepsilon^\dagger(\boldsymbol{p}) \partial_{p_\mu} U_\varepsilon(\boldsymbol{p}) = 0$. Plugging this into Eq. (S14) takes us to

$$\begin{aligned} G_{\mu\nu}^\varepsilon(\boldsymbol{p}) &= \frac{1}{2} \text{Tr} \left[ \partial_{p_\mu} U_\varepsilon^\dagger(\boldsymbol{p}) U_\varepsilon(\boldsymbol{p}) U_\varepsilon^\dagger(\boldsymbol{p}) \partial_{p_\nu} U_\varepsilon(\boldsymbol{p}) \right] \\ &= \frac{1}{2} \text{Tr} \left[ U_\varepsilon^\dagger(\boldsymbol{p}) \partial_{p_\mu} U_\varepsilon(\boldsymbol{p}) \partial_{p_\nu} U_\varepsilon^\dagger(\boldsymbol{p}) U_\varepsilon(\boldsymbol{p}) \right] \\ &= \frac{1}{2} \text{Tr} \left[ \partial_{p_\mu} U_\varepsilon(\boldsymbol{p}) \partial_{p_\nu} U_\varepsilon^\dagger(\boldsymbol{p}) U_\varepsilon(\boldsymbol{p}) U_\varepsilon^\dagger(\boldsymbol{p}) \right] \\ &= \frac{1}{2} \text{Tr} \left[ \partial_{p_\mu} U_\varepsilon(\boldsymbol{p}) \partial_{p_\nu} U_\varepsilon^\dagger(\boldsymbol{p}) \right]. \end{aligned}$$ (S15)

From Eq. (S15), we find $G_{\mu\nu}^\varepsilon = (G_{\mu\nu}^\varepsilon)^*$. It is easy to see that $G_{\mu\nu}^\varepsilon = G_{\nu\mu}^\varepsilon$, since the trace is invariant under cyclic permutations $\text{Tr}(AB) = \text{Tr}(BA)$ for arbitrary matrices $A$ and $B$. In conclusion, the FQMT $G_{\mu\nu}^\varepsilon$ is a real symmetric tensor.

## S-III. PROOF OF EQ. (9) IN THE MAIN TEXT

We recall that the winding number and the FQMT are respectively given by

$$W = \frac{1}{2\pi^2} \int_{\mathbb{T}^3} d^3\boldsymbol{p} \ \epsilon^{\mu\nu\rho\lambda} d_\mu^\varepsilon \partial_{k_x} d_\nu^\varepsilon \partial_{k_y} d_\rho^\varepsilon \partial_t d_\lambda^\varepsilon,$$ (S16)

$$G_{\mu\nu}^\varepsilon = \sum_{i=0}^3 \partial_{p_\mu} d_i^\varepsilon \partial_{p_\nu} d_i^\varepsilon.$$ (S17)

Let us now express the FQMT in the following matrix form,

$$\begin{aligned} G^\varepsilon &= \begin{bmatrix} G_{xx}^\varepsilon & G_{xy}^\varepsilon & G_{xt}^\varepsilon \\ G_{yx}^\varepsilon & G_{yy}^\varepsilon & G_{yt}^\varepsilon \\ G_{tx}^\varepsilon & G_{ty}^\varepsilon & G_{tt}^\varepsilon \end{bmatrix} \\ &= \sum_i \begin{bmatrix} \partial_{k_x} d_i^\varepsilon \partial_{k_x} d_i^\varepsilon & \partial_{k_x} d_i^\varepsilon \partial_{k_y} d_i^\varepsilon & \partial_{k_x} d_i^\varepsilon \partial_t d_i^\varepsilon \\ \partial_{k_y} d_i^\varepsilon \partial_{k_x} d_i^\varepsilon & \partial_{k_y} d_i^\varepsilon \partial_{k_y} d_i^\varepsilon & \partial_{k_y} d_i^\varepsilon \partial_t d_i^\varepsilon \\ \partial_t d_i^\varepsilon \partial_{k_x} d_i^\varepsilon & \partial_t d_i^\varepsilon \partial_{k_y} d_i^\varepsilon & \partial_t d_i^\varepsilon \partial_t d_i^\varepsilon \end{bmatrix}. \end{aligned}$$ (S18)

The determinant of $G^\varepsilon$ is equal to the determinant of the following matrix, say $G^{\varepsilon'}$:

$$G^{\varepsilon'} = \begin{bmatrix} 1 & 0 & 0 & 0 \\ 0 & G_{xx}^\varepsilon & G_{xy}^\varepsilon & G_{xt}^\varepsilon \\ 0 & G_{yx}^\varepsilon & G_{yy}^\varepsilon & G_{yt}^\varepsilon \\ 0 & G_{tx}^\varepsilon & G_{ty}^\varepsilon & G_{tt}^\varepsilon \end{bmatrix}.$$ (S19)

Then we can decompose $G^{\varepsilon'}$ as $G^{\varepsilon'} = AA^T$, with

$$A = \begin{bmatrix} d_0^\varepsilon & d_1^\varepsilon & d_2^\varepsilon & d_3^\varepsilon \\ \partial_{k_x} d_0^\varepsilon & \partial_{k_x} d_1^\varepsilon & \partial_{k_x} d_2^\varepsilon & \partial_{k_x} d_3^\varepsilon \\ \partial_{k_y} d_0^\varepsilon & \partial_{k_y} d_1^\varepsilon & \partial_{k_y} d_2^\varepsilon & \partial_{k_y} d_3^\varepsilon \\ \partial_t d_0^\varepsilon & \partial_t d_1^\varepsilon & \partial_t d_2^\varepsilon & \partial_t d_3^\varepsilon \end{bmatrix},$$ (S20)

where the relation $\sum_i (d_i^\varepsilon)^2 = 1$ is used [3]. The determinant of the matrix A is

$$\det A = \det A^T = \epsilon^{\mu\nu\rho\lambda} d_\mu^\varepsilon \partial_{k_x} d_\nu^\varepsilon \partial_{k_y} d_\rho^\varepsilon \partial_t d_\lambda^\varepsilon. \tag{S21}$$

Therefore, we have

$$\sqrt{\det G^\varepsilon} = |\epsilon^{\mu\nu\rho\lambda} d_\mu^\varepsilon \partial_{k_x} d_\nu^\varepsilon \partial_{k_y} d_\rho^\varepsilon \partial_t d_\lambda^\varepsilon|. \tag{S22}$$

Then we conclude with the topological bound for the FQV: vol $\geq 2\pi^2|W|$. We note that this relation can also be proved using more general state evolution, see Sec. S-VII for details.

## S-IV. EDGE STATES AND BAND INVERSIONS IN THE FLOQUET CHERN INSULATOR

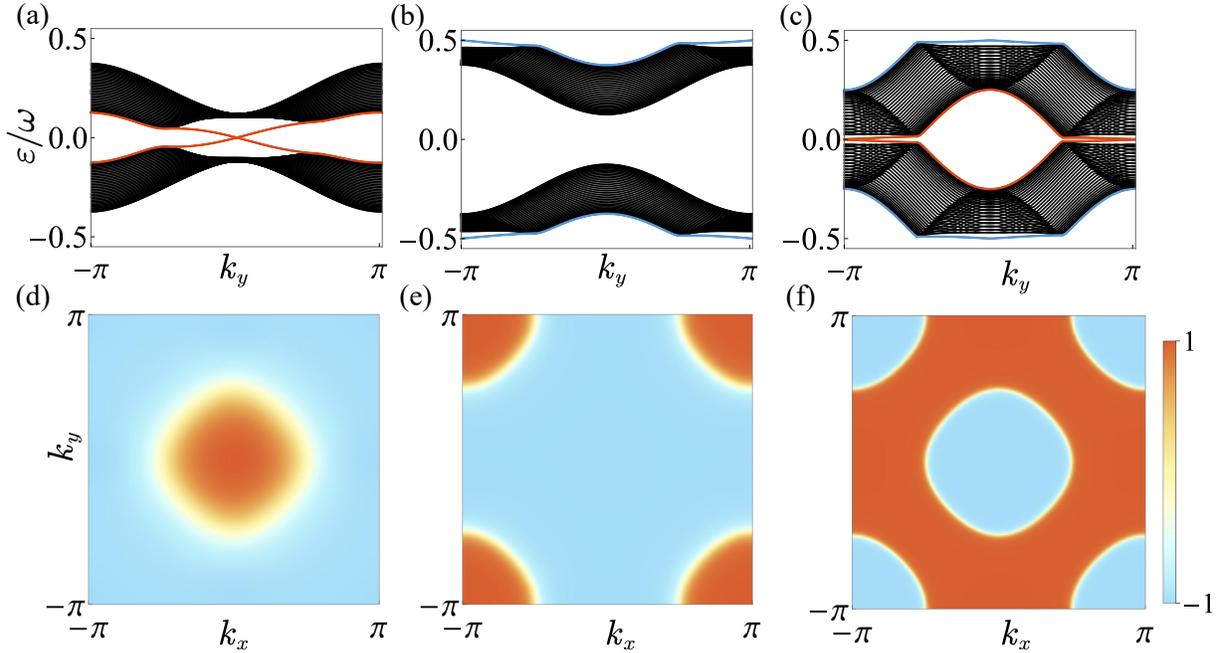

FIG. S1. Up panel: Quasienergy spectrum of the Floquet Qi-Wu-Zhang model with system length $L_x = 40$ under open boundary condition imposed along the $x$ direction, in (a) the topological 0 phase with $m = 5$ and $\omega = 8$, (b) the topological $\pi$ phase $\pi$ phase $m = 15$ and $\omega = 8$, and (c) topological 0-$\pi$ phase $m = 15$ and $\omega = 4$.. Down panel: The polarization of the lower Floquet eigenstates in (d) the topological 0 phase with $m = 5$ and $\omega = 8$, (e) the topological $\pi$ phase $\pi$ phase $m = 15$ and $\omega = 8$, and (f) topological 0-$\pi$ phase $m = 15$ and $\omega = 4$. The parameters used here are the same with Fig. 1 in the main text.

In the Floquet topological phases, the system acquires topology-protected in-gap modes at the edges. In Figs. S1(a)-S1(c), we plot the quasienergy spectrum for the Floquet Qi-Wu-Zhang model with cylindrical geometry. We can see that the gapless edge states can show up in the 0 gap or $\pi$ gap or both according to the type of the topological phases.

The topology can also be reflected in the band inversion process of the bulk bands. The bulk gap closes and reopens when the topological phase transition occurs. Typically, this process is accompanied with band inversions. For the two-band model $H_1(\boldsymbol{k}, t) = \boldsymbol{h}(\boldsymbol{k}, t) \cdot \boldsymbol{\sigma}$, with $\boldsymbol{h}(\boldsymbol{k}, t) = (v_x \sin k_x, v_y \sin k_y, M(t) - v_z \cos k_x - v_z \cos k_y)$, the first two terms in the Hamiltonian can be regarded as spin-orbit coupling. The spin-orbit coupling opens a topological gap at the band bottom and band top, where the band inversion occurs. The feature of band inversions can be revealed by the polarization of the lower Floquet eigenstate: $P = \langle u_- | \sigma_z | u_- \rangle$. In Fig. S1, we plot the polarization for distinct topological phases. The polarization switches sign across the band inversion lines, implying its topological nature.

## S-V.   THE DISTRIBUTION OF THE FLOQUET QUANTUM METRIC

For reference, we show all the nine components of the Floquet quantum metric $G^0_{\mu\nu}(t = \frac{T_1}{2})$ and $G^\pi(t = \frac{T_1}{2})$ for the Floquet Qi-Wu-Zhang model in Figs. S2 and S3, respectively. The system is in the topological 0 phase for the parameters chosen here. It is easy to see that they are symmetric, $G^\epsilon_{\mu\nu} = G^\epsilon_{\nu\mu}$. The off-diagonal terms can be negative such as $G_{yx}$, $G_{tx}$ and $G_{ty}$, which also happens for some static systems. However, the distance $D^2$ must be positive semidefinite as guaranteed by its definition.

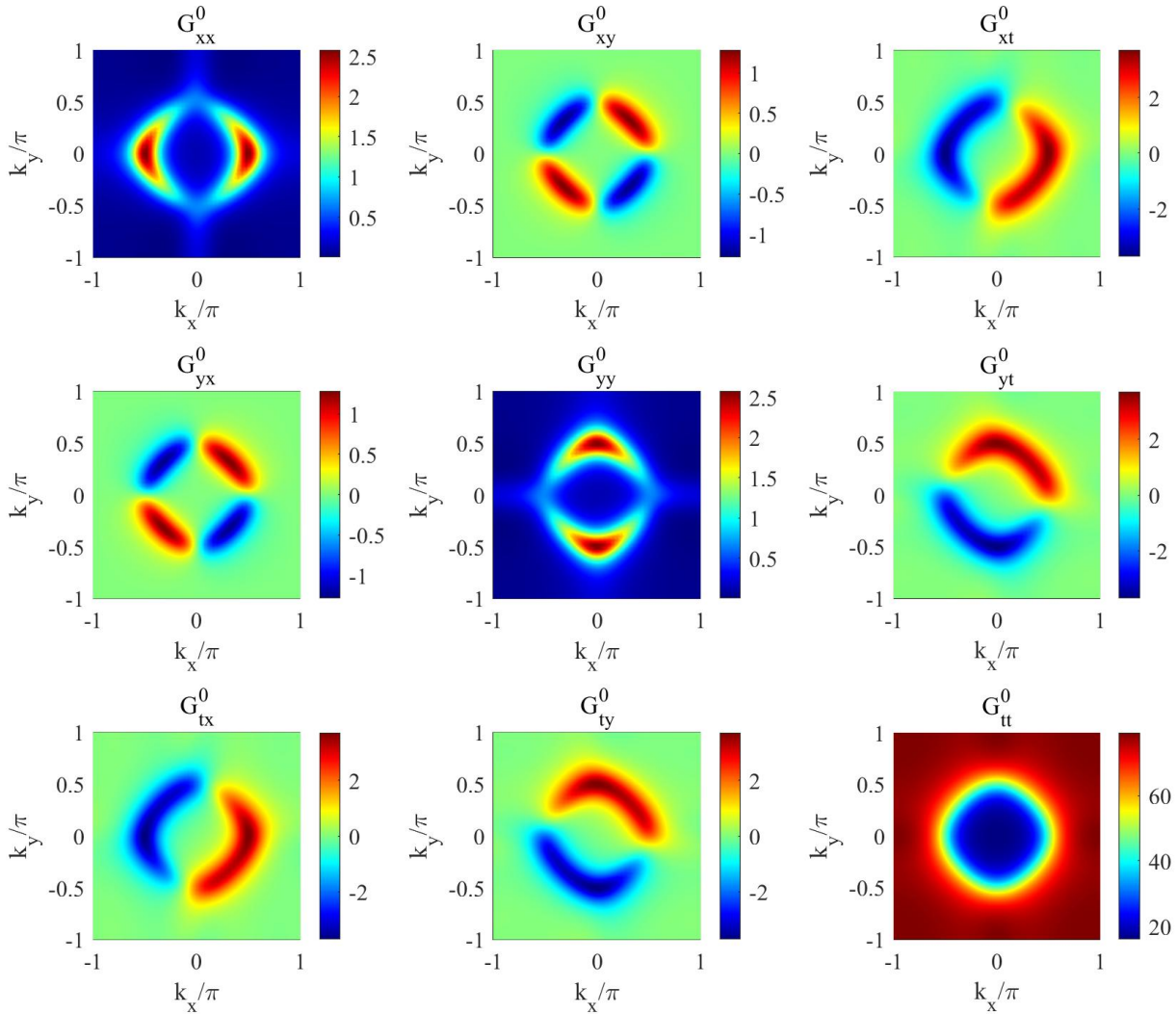

FIG. S2. The components of the Floquet quantum metric $G^0_{\mu\nu}(t = \frac{T_1}{2})$ for the Floquet Qi-Wu-Zhang model, with $m = 5$, $\omega = 8$, $T_1 = 0.6T$, $T_2 = 0.4T$.

## S-VI.   FLOQUET QUANTUM METRIC TENSOR IN THE HIGH-FREQUENCY LIMIT

In the high-frequency limit, the system can be described by a time-independent effective Hamiltonian, as is well known from the Magnus expansion in the order of $1/\omega$ [4]. For simplicity, we consider an effective Hamiltonian of the form [5]

$$H_{\text{eff}}(\boldsymbol{k}) = -\frac{2\pi}{T} P(\boldsymbol{k}), \tag{S23}$$

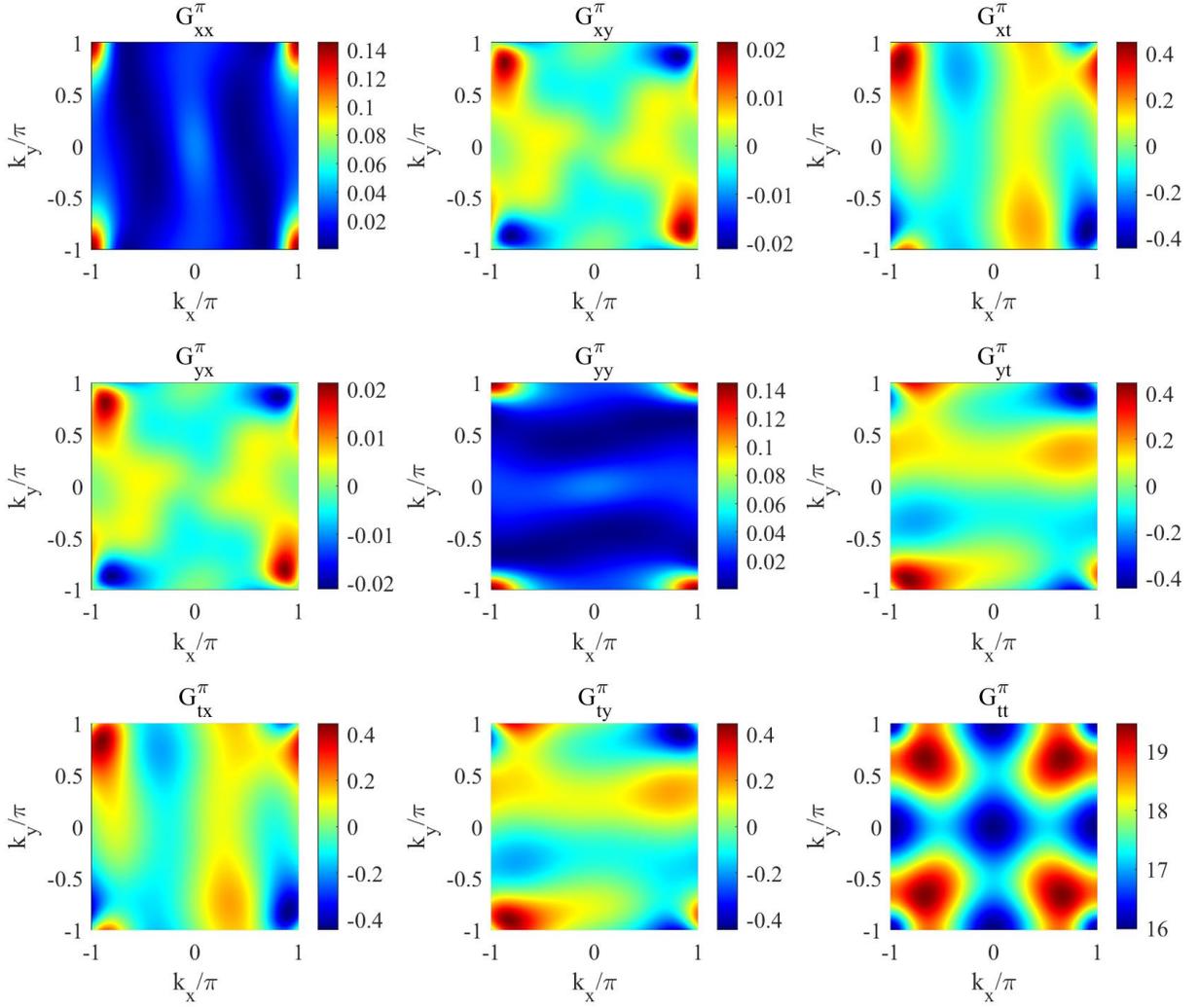

FIG. S3. The components of the Floquet quantum metric $G_{\mu\nu}^{\pi}(t = \frac{T_1}{2})$ for the Floquet Qi-Wu-Zhang model, with $m = 5$, $\omega = 8$, $T_1 = 0.6T$, $T_2 = 0.4T$.

where $P(\boldsymbol{k})$ is a projection operator defined above. The time evolution operator reads

$$
\begin{aligned}
U &= (e^{2\pi it/T} - 1)P + 1, \\
U^{-1} &= (e^{-2\pi it/T} - 1)P + 1.
\end{aligned}
\tag{S24}
$$

Next we consider different matrix elements of the FQMT based on $U$, according to our general treatment elaborated in the main text. First of all,

$$
\begin{aligned}
G_{k_i k_j} &= \frac{1}{2}\operatorname{Tr}[\partial_{k_i} U^{\dagger} \partial_{k_j} U] = \left[1 - \cos\left(\frac{2\pi t}{T}\right)\right] \operatorname{Tr}[\partial_{k_i} P \partial_{k_j} P], \\
G_{t k_i} &= \frac{1}{2}\operatorname{Tr}[\partial_t U^{\dagger} \partial_{k_i} U] = \frac{2\pi i}{T}\left[1 - \cos\left(\frac{2\pi t}{T}\right)\right] \operatorname{Tr}[P \partial_{k_i} P], \\
G_{tt} &= \frac{1}{2}\operatorname{Tr}[\partial_t U^{\dagger} \partial_t U] = \frac{2\pi^2}{T^2}\operatorname{Tr}[P] = \frac{2\pi^2}{T^2}.
\end{aligned}
\tag{S25}
$$

Interestingly, the time average of $G_{k_i k_j}$ reduces to the conventional quantum metric,

$$
\overline{G}_{k_i k_j} \equiv \int \frac{dt}{T} G_{k_i k_j} = \int \frac{dt}{T}\frac{1}{2}\operatorname{Tr}[\partial_{k_i} U^{\dagger} \partial_{k_j} U] = \operatorname{Tr}[\partial_{k_i} P \partial_{k_j} P],
\tag{S26}
$$

because $g_{k_i k_j} = \text{Tr}[\partial_{k_i} P \partial_{k_j} P]$.

Regarding other matrix elements of the FQMT, one quickly has $G_{t k_j} = 0$. Due to $P^2 = P$, we have $\partial_{k_i} P^2 = (\partial_{k_i} P) P + P(\partial_{k_i} P) = \partial_{k_i} P$. Then we have $2 \text{Tr}[P \partial_{k_i} P] = \text{Tr}[\partial_{k_i} P] = \partial_{k_i} \text{Tr}[P] = 0$. Therefore, $G_{t k_j} = \frac{1}{2} \text{Tr}[\partial_t U^\dagger \partial_{k_i} U] = 0$. Now the Floquet quantum volume is

$$
\begin{aligned}
\text{vol} &= \int d^3 \boldsymbol{p} \sqrt{\det(G_{\mu\nu})} \\
&= \frac{\sqrt{2}\pi}{T} \int dk_x dk_y \int dt [1 - \cos\left(\frac{2\pi t}{T}\right)] \sqrt{\text{Tr}[\partial_{k_x} P \partial_{k_x} P] \cdot \text{Tr}[\partial_{k_y} P \partial_{k_y} P] - \text{Tr}[\partial_{k_x} P \partial_{k_y} P] \cdot \text{Tr}[\partial_{k_y} P \partial_{k_x} P]} \\
&= \sqrt{2}\pi \int dk_x dk_y \sqrt{g_{k_\mu k_\nu}} \\
&= \sqrt{2}\pi \text{vol}_{\text{static}}.
\end{aligned}
\tag{S27}
$$

Here $\text{vol}_{\text{static}}$ is the quantum volume in static systems.

Furthermore, it has been shown that for such Hamiltonian the winding number for $U$ reduces to the Chern number [5],

$$
W = \frac{1}{2\pi i} \int dk_x dk_y \text{Tr}(P[\partial_{k_x} P, \partial_{k_y} P]),
\tag{S28}
$$

with the integrand just the Berry curvature.

### S-VII. EXPRESSING THE FLOQUET QUANTUM METRIC TENSOR WITH STATE EVOLUTION

The matrix elements of the evolution operator $U$ can be found by evolving an initial state $\psi_a$ (which is assumed to be $\boldsymbol{k}$-independent) $U_{ab} = \langle \psi_a | U | \psi_b \rangle$ with indices $a, b = 1, 2$ for a two-band model. By denoting the time-evolved state as $\Psi_a = U|\psi_a\rangle$, we have

$$
\begin{aligned}
G_{\mu\nu} &= \frac{1}{2} \text{Tr}[\partial_{p_\mu} U^\dagger \partial_{p_\nu} U] \\
&= \frac{1}{2} \sum_{a,b} \partial_{p_\mu} U^\dagger_{ab} \partial_{p_\nu} U_{ba} \\
&= \frac{1}{2} \sum_{a,b} \partial_{p_\mu} \langle \Psi_a | \psi_b \rangle \partial_{p_\nu} \langle \psi_b | \Psi_a \rangle \\
&= \frac{1}{2} \sum_a \langle \partial_{p_\mu} \Psi_a | \partial_{p_\nu} \Psi_a \rangle,
\end{aligned}
\tag{S29}
$$

where we have used $\sum_b |\psi_b\rangle\langle\psi_b| = \mathbb{I}$. The general form of $\Psi_a(\boldsymbol{k}, t)$ can be written in terms of two complex scalar fields $z_1(\boldsymbol{k}, t)$ and $z_2(\boldsymbol{k}, t)$ as $\Psi_1 = (z_1, z_2^*)^T$ and $\Psi_2 = (-z_2, z_1^*)^T$, then

$$
\begin{aligned}
G_{\mu\nu} &= \frac{1}{2}(\partial_{p_\mu} \Psi_1^\dagger \partial_{p_\nu} \Psi_1 + \partial_{p_\mu} \Psi_2^\dagger \partial_{p_\nu} \Psi_2) \\
&= \frac{1}{2} \sum_i (\partial_{p_\mu} z_i^* \partial_{p_\nu} z_i + \partial_{p_\mu} z_i \partial_{p_\nu} z_i^*).
\end{aligned}
\tag{S30}
$$

Furthermore, it has been shown that the winding number Eq. (S16) can be reduced to a Hopf number [6],

$$
\begin{aligned}
W &= -\frac{1}{4\pi^2} \int dp^3 \epsilon_{ijk} \Psi_1^\dagger \partial_i \Psi_1 \partial_j \Psi_1^\dagger \partial_k \Psi_1 \\
&= -\frac{1}{4\pi^2} \int dp^3 \epsilon_{ijk} (z_1^* \partial_i z_1 \partial_j z_2^* \partial_k z_2 + z_2^* \partial_i z_2 \partial_j z_1^* \partial_k z_1).
\end{aligned}
\tag{S31}
$$

Notably, we can also prove the bound for the Floquet quantum volume using Eq. (S30) and Eq. (S31). First, let us denote $z_1 = x_1 + iy_1$ and $z_2 = x_2 + iy_2$, with $x_1^2 + y_1^2 + x_2^2 + y_2^2 = 1$. Then we define $\phi = (x_1, y_1, x_2, y_2)$. Now we can rewrite the metric tensor as,

$$
G_{\mu\nu} = \sum_{a=1}^4 \partial_{p_\mu} \phi_a \partial_{p_\nu} \phi_a.
\tag{S32}
$$

On the other hand, we can rewrite the Hopf number as

$$W = \frac{1}{12\pi^2} \int d^3p \; \epsilon_{abcd} \epsilon^{ijk} \phi^a \partial_i \phi^b \partial_j \phi^c \partial_k \phi^d$$
$$= \frac{1}{2\pi^2} \int d^3p \; \epsilon_{abcd} \epsilon^{ijk} \phi^a \partial_i \phi^b \partial_j \phi^c \partial_k \phi^d. \tag{S33}$$

Therefore, similar to Sec. III, we conclude that vol $\geq 2\pi^2 |W|$.

## S-VIII. PHASES WITH HIGHER WINDING NUMBERS FOR THE FLOQUET SSH MODEL

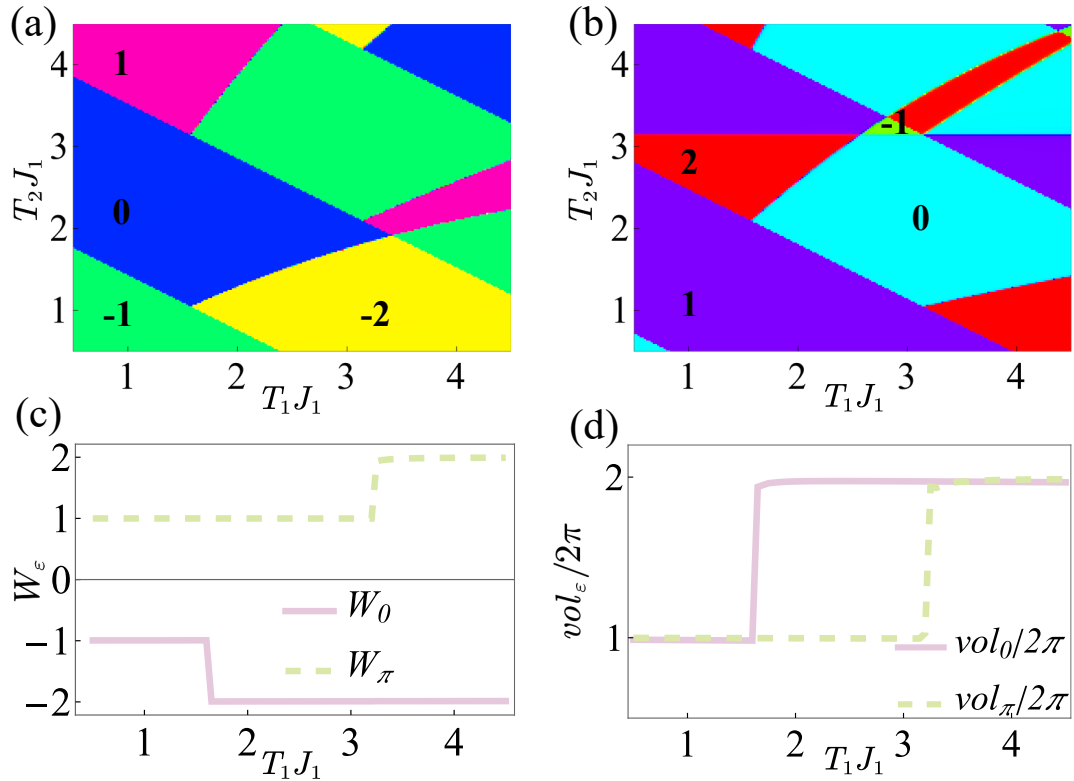

FIG. S4. (a) Phase diagram for the Floquet SSH model characterized by $W_0$. (b) Phase diagram for the Floquet SSH model characterized by $W_\pi$. Other parameters are chosen as $J/J_1 = 1$ and $q = 2$. (c) The winding numbers at the section of $T_2 J_1 = 1$. (d) The Floquet quantum volume as the function of $T_1 J_1$.

We map out the phase diagram for the Floquet SSH model on the $T_1$-$T_2$ plane in Figs. S4(a) and S4(b). Rich topological phases are induced by changing the driving parameters. There are phases with higher winding numbers $|W_{0,\pi}| > 1$ thus carrying multiple edge channels. The presence of such phases can be understood by the periodic driving inducing effective long-range hoppings. From the topological bound we proved in the main text: vol$_\varepsilon \geq 2\pi|W_\varepsilon|$, we expect that the phases with higher winding numbers have larger quantum volume. In Fig. S4(c) and S4(d), we present the change of quantum volume as the topological phase transition occurs. We clearly see that the quantum volume increases when the system enters the phase with $|W| = 2$.

## S-IX. THE QUANTUM VOLUME OF THE MOMENTUM-TIME SPACE FOR THE FLOQUET SSH MODEL

In this subsection, we investigate the full FQMT in the momentum-time space, without adopting the reduction done in the main text. In Fig. S5, we present our computational results of the distribution of the Floquet quantum

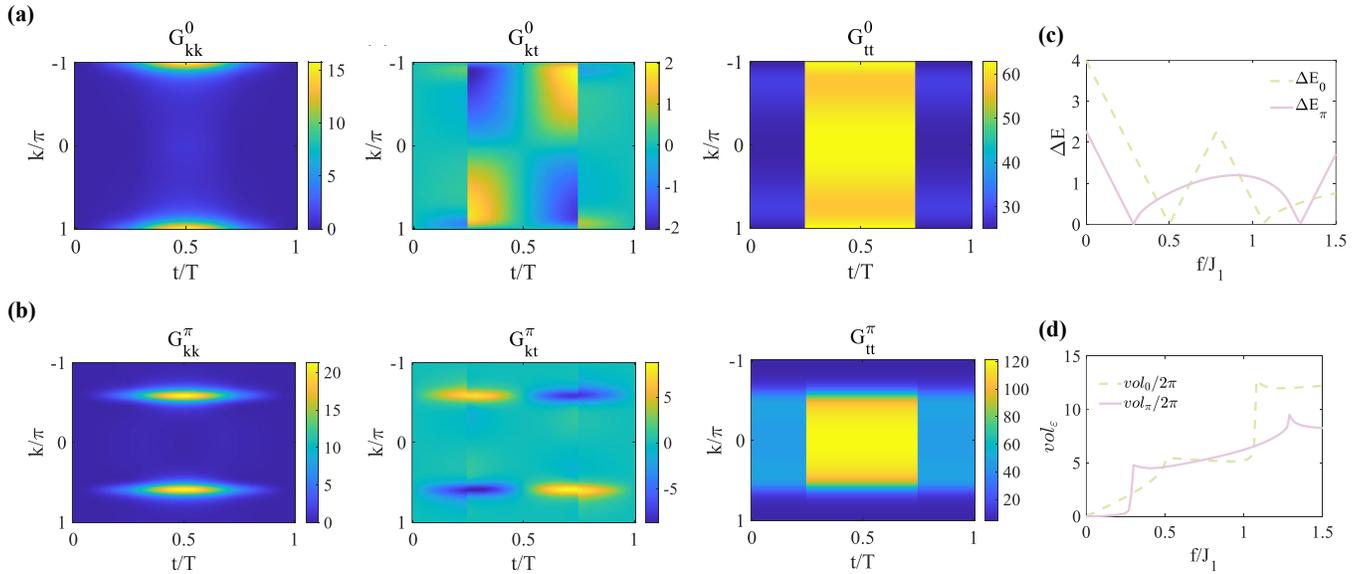

FIG. S5. (a) The components of the Floquet quantum metric $G_{\mu\nu}^0$ for the Floquet SSH model . (b) The components of the Floquet quantum metric $G_{\mu\nu}^\pi$ for the Floquet SSH model. Here $\mu, \nu = k, t$. (c) The quasienergy gap for Floquet SSH model as the function of $f/J_1$. (d) The Floquet quantum volume of momentum-time space as the function of $f/J_1$. The parameters we use is $q = 3$, $\omega = 2\pi$, $T_1 = T_2$, and $f/J_1 = 0.75$ for (a) and (b).

metric in the momentum-time space. Then we calculate the associated quantum volume of the momentum-time space as follows:

$$vol_\varepsilon = \oint dt dk \sqrt{G_{tt}^\varepsilon G_{kk}^\varepsilon - G_{kt}^\varepsilon G_{tk}^\varepsilon}. \tag{S34}$$

In Fig. S5(a) and S5(b), we find that $G_{tt}$ is almost uniformly distributed in each driving step, and $G_{tt} \gg G_{kt} = G_{tk}$. Furthermore, $G_{kk}$ slowly varies as time evolves. This implies that quantum volume of the momentum-time space is dominated by the integral of $\sqrt{G_{kk}}$, $vol_\varepsilon \sim \alpha \oint dt dk \sqrt{G_{kk}}$, where $\alpha$ is determined by the integral of $\sqrt{G_{tt}}$. As a result, we expect that the quantum volume of the momentum-time space will generate similar profile with that of the momentum space alone, a reduction adopted in the main text based on symmetry considerations. Therefore, the quantum volume of the momentum space can well capture the main feature of this system due to the symmetry restriction.

Furthermore, we find that the system will exhibit a large FQV in a trivial phase, as shown in Fig. S5(d) here and Fig. 3(b) in the main text. This might be attributed to the small gap in this phase, see S5(c).

Since the topological invariant for the chiral class does not involve the time integral, one may wonder if the topological invariants and quantum metric can be defined merely based on the effective Hamiltonian defined in the main text. To see this, let us consider another way in finding the winding numbers for the 0 gap and $\pi$ gap, using two symmetric frames [7]. Using that the Floquet system has a degree of freedom on the Floquet gauge, one can choose two different starting times and then obtain two symmetric evolution operators,

$$U_{T,1} = e^{-iH_1 T_1/2} e^{-iH_2 T_2} e^{-iH_1 T_1/2},$$
$$U_{T,2} = e^{-iH_2 T_2/2} e^{-iH_1 T_1} e^{-iH_2 T_2/2}. \tag{S35}$$

From this procedure one can now define two effective Hamiltonians $H_1$ and $H_2$ from $U_{T,1}$ and $U_{T,2}$. Interestingly, the winding numbers for the 0 gap and $\pi$ gap can be extracted from the winding numbers $W_1$ and $W_2$ for $H_1$ and $H_2$,

$$W_0 = (W_1 + W_2)/2, \quad W_\pi = (W_1 - W_2)/2. \tag{S36}$$

From the equation above, it is now clear that the winding numbers still cannot be obtained using one single effective Hamiltonian. Therefore, from this angle one can also conclude that it will be improper to use one single effective Hamiltonian to link quantum geometry with the Floquet band topology. We hence do not expect any means to reduce the FQMT more than the reduction discussed in the main text for class AIII.



\end{document}